# Bayesian Active Learning for Scanning Probe Microscopy: from Gaussian Processes to Hypothesis Learning

Maxim Ziatdinov,[1,2,*] Yongtao Liu,[2] Kyle Kelley,[2] Rama Vasudevan,[2] and Sergei V. Kalinin[3,*]

[1] Computational Sciences and Engineering Division and [2] Center for Nanophase Materials Sciences, Oak Ridge National Laboratory, Oak Ridge, TN 37831, USA
[3] Department of Materials Sciences and Engineering, University of Tennessee, Knoxville, TN 37996, USA

**Abstract**

Recent progress in machine learning methods, and the emerging availability of programmable interfaces for scanning probe microscopes (SPMs), have propelled automated and autonomous microscopies to the forefront of attention of the scientific community. However, enabling automated microscopy requires the development of task-specific machine learning methods, understanding the interplay between physics discovery and machine learning, and fully defined discovery workflows. This, in turn, requires balancing the physical intuition and prior knowledge of the domain scientist with rewards that define experimental goals and machine learning algorithms that can translate these to specific experimental protocols. Here, we discuss the basic principles of Bayesian active learning and illustrate its applications for SPM. We progress from the Gaussian Process as a simple data-driven method and Bayesian inference for physical models as an extension of physics-based functional fits to more complex deep kernel learning methods, structured Gaussian Processes, and hypothesis learning. These frameworks allow for the use of prior data, the discovery of specific functionalities as encoded in spectral data, and exploration of physical laws manifesting during the experiment. The discussed framework can be universally applied to all techniques combining imaging and spectroscopy, SPM methods, nanoindentation, electron microscopy and spectroscopy, and chemical imaging methods, and can be particularly impactful for destructive or irreversible measurements.

**Corresponding emails:** ziatdinovma@ornl.gov; sergei2@utk.edu

Scanning probe microscopy (SPM) has become a mainstay of multiple scientific fields including materials science, condensed matter physics, biology, and medicine.[1-3] The relatively low cost and ease of use of ambient and environmental SPM systems made them ubiquitous across academic and industrial labs, with nanometer resolution images of the surface topographies,[4] electric,[5,6] magnetic,[7,8] and mechanical[9,10] properties becoming commonplace. Liquid SPMs provide deep insights into the electrochemical processes[11,12] or dynamics and mechanical properties of biological objects *in vivo*. Finally, the ultra-high vacuum SPMs, typically operated as a part of the surface science/deposition systems, allow exploring structures and electronic properties of surfaces on atomic level,[13,14] and are now becoming the key component of the development of enabling technologies and basic knowledge in quantum sciences.[15]

Beyond structural and functional imaging, all SPM methods can modify the state of the surface. The trivial (and undesired) example of this is the surface damage during scanning, often associated with tip degradation. Under controlled conditions, the tip-surface interactions can be used for mechanical and electrochemical lithography,[16,17] altering surface morphology or inducing local electrochemical transformations. In ferroelectric materials, the bias applied to the probe can be used to change the polarization state of the surface, modifying and creating domain patterns.[18,19] Finally, the probe can be used to move individual atoms and molecules along the surface, inducing the chemical reactions and assembling structures from individual atoms.[20,21]

This broad variety of SPM methods was established in late 1990s, and progress since then has been primarily associated with the improvements in instrument design, the introduction of a broad range of environmental capabilities, and in several cases, with the increase of the volumes of the acquired data.[22] For the latter, examples include 'hyperspectral' imaging methods, where the spectroscopic data sets are acquired over the rectangular grid of points, giving rise to 3D data sets. The well-known examples of these methods are continuous imaging tunneling spectroscopy (CITS)[23] in scanning tunneling microscopy and force-volume imaging[24] in atomic force microscopy. A particularly broad range of spectroscopic methods have been developed in the context of Piezoresponse Force Microscopy (PFM),[25] this development made possible by the intrinsically quantitative nature of PFM and the strong connection between the measured signal and the fundamental physics of the electric polarization order parameter.[26,27]

The acquisition of large data volumes has further required the development of methods for the analysis and interpretation of these data sets. In cases for which physical models are known, individual spectra can be converted to relevant physical descriptors. For example, force-distance curves can be analyzed using continuum mechanical models to yield the indentation modulus,[28] whereas amplitude-frequency curves in band excitation can be fitted to yield resonance frequency and response amplitude.[29] However, in most cases the models are not available or poorly known, necessitating the development of machine learning methods for data reduction, with examples ranging from principal component analysis[30] to neural network recognition[31] and manifold learning.[32]

The rapid growth in deep learning methods in computer vision, robotics, and medical image analysis over the last decade has further resulted in applications of these methods within SPM imaging. Here, deep learning methods have been used for identification of atoms and defects on a Si surface,[33] recognition of DNA molecules,[34] imaging biological objects, identifying ferroelectric domain walls,[35] and conditioning the state of the probe[36, 37].

Finally, the availability of multiple information channels representing the structure and physical functionalities in SPM have further enabled structure-property data mining. Here, the relationships between the individual structural elements and functionalities determined via spectroscopic methods are discovered. Examples of such approaches include the relationship between ferroelectric domain structures and local hysteresis loops in PFM[38], and local electronic properties and structure in graphene.[39]

Most applications of machine learning methods in SPM have been realized for already acquired data, effectively exploring the internal correlations between spatial and spectroscopic dimensions of a static dataset that has been acquired over a rectangular grid of points. At the same time, many commercial instruments allow (or can be modified to allow) some degree of external control, moving the probe to predefined locations, initiating acquisition of spectroscopic data, and generating custom probe signals. While early reports on controlled experimental path have been available for decades,[40-42] the recent introduction of programmable controllers made these approaches considerably more accessible. This have generated much enthusiasm on combining the machine-learning based data analysis within the microscope data acquisition loop, i.e., the concept of automated and autonomous microscopy. Correspondingly, a number of opinion and analysis pieces have been published over the last few years.[33, 36, 37, 43-46]

The analysis of existing machine learning algorithms and frameworks suggest that the analysis of static data after the acquisition and active learning during the experiment require considerably different methods. Analysis of static data sets can be performed using an immense variety of supervised, semi-supervised, and unsupervised methods towards the goals of semantic segmentation, regression, clustering, etc. These methods allow researchers to explore internal data structure, establish correlations between data and labels, or establish the causal links between multiple data channels. However, many of these methods will perform poorly in dynamic environments due to out-of-distribution shifts. In the context of microscopy, a deep learning network trained on one data set will generally perform poorly on the data set obtained on the same material under different imaging conditions. Perhaps even more fundamentally, supervised ML methods require fully labeled data sets for analysis, limiting the range of experimental tasks they can be applied to.

Recently, we have critically explored the potential of machine learning methods for automated microscopies.[44] Here, we illustrate these approaches for SPM (Figure 1) and demonstrate the development of ML-based experimental workflows based on Bayesian methods. We discuss the AE in microscopy based on the (explicit or implicit) prior knowledge involved in the experiment design, the knowledge used during the experiment, and the goal of the experiment (reward). We note that all three components including the reward should be established explicitly prior to the initial definition of the experimental workflow, building the required ML components, testing on the pre-acquired "ground truth" data, and deployments on an operational microscope.

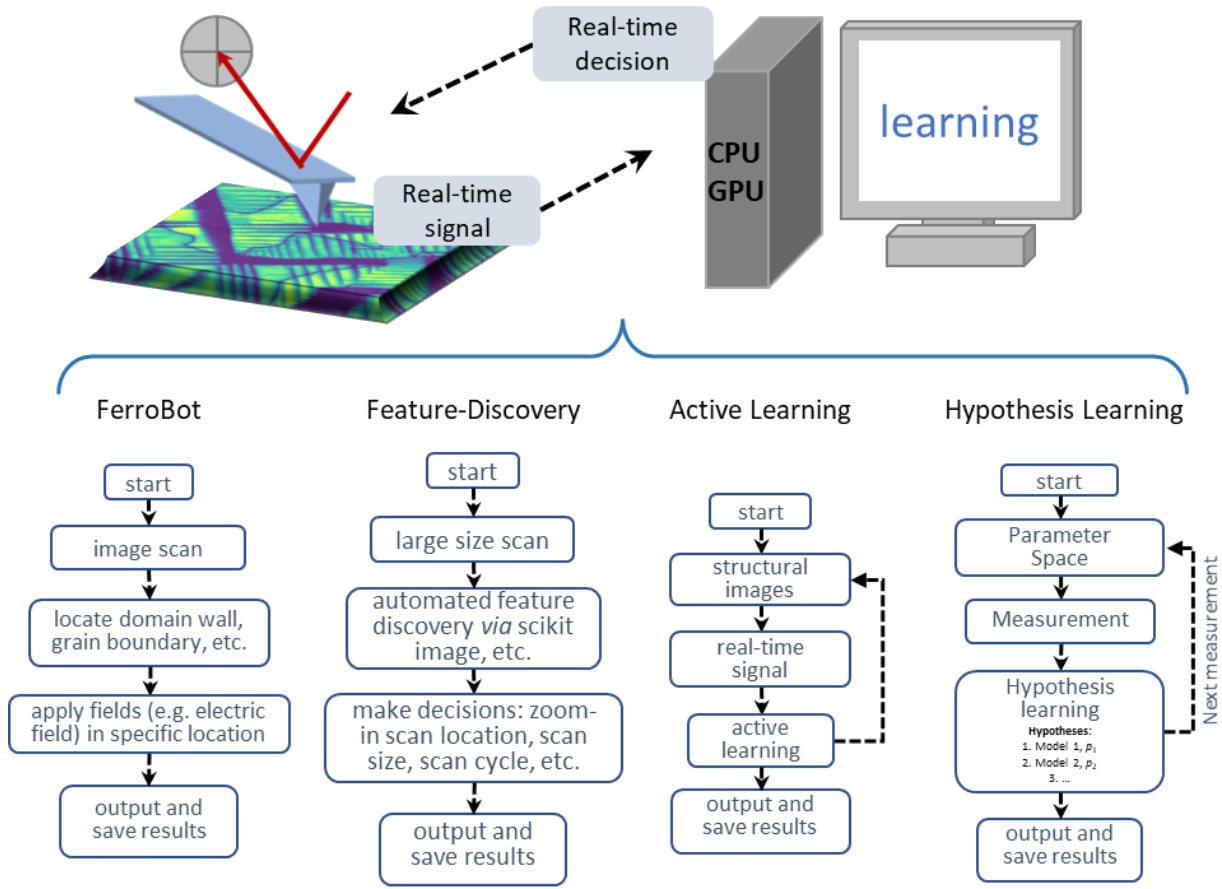

**Figure 1.** Machine learning workflows designed for automated SPM. FerroBOT and Feature-discovery algorithms are based on decision making, e.g., application of bias pulse or spectroscopic measurement, based on predefined rules on a line by line or image by image basis. Comparatively, in active learning the operators define what aspects of the measured spectral signal are of interest and the algorithms discovers what microstructural elements exhibit this behavior. Finally, in hypothesis learning, the algorithm optimizes the function of interest over a certain parameter space if physical model of this behavior (structured Gaussian Process) or multiple competing models (hypothesis learning) models of this behavior are available. Such examples can include optimization of microscope resolution as a function of tuning parameters, or exploring the growth mechanism of ferroelectric domains as a function of probe bias pulse amplitude and pulse width.

**Results and Discussion**

1. **Why automated experiment**

To illustrate the concepts that emerge in the context of the automated experiment, let us explore a particular experimental workflow utilizing the example of Piezoresponse Force Microscopy (PFM) imaging and spectroscopy. However, we note that the logic espoused here applies to any other probe method including other SPM modalities, scanning transmission electron microscopy (STEM), or optical and mass-spectrometric chemical imaging methods.

PFM is a functional SPM mode for probing local electromechanical properties of piezoelectric and ferroelectric materials. In a PFM measurement, a conductive probe tip is brought into contact with a sample and an AC bias is applied to the tip, resulting in local expansion or contraction of the sample. This interaction force between tip and sample is due to the converse piezoelectric effect, and is detected by an optical beam photodiode system. As such, the sample's piezoelectric response is measured as an amplitude and phase signal corresponding to the polarization magnitude and orientation, respectively. PFM imaging mode is when the tip is scanned across the sample, and the piezoelectric response (piezoresponse) at each pixel is simultaneously recorded. This mode allows mapping of ferroelectric domain structures. PFM switching spectroscopy mode—in which the AC bias is superimposed onto a triangular sweep DC bias waveform—allows measuring the piezoresponse as a function of DC bias, namely, piezoresponse-voltage hysteresis loops, at each pixel of an image. Parameters encoded in the piezoresponse-voltage hysteresis loops, e.g. coercive field, nucleation bias, loop area/width, and remnant polarization, can be extracted and plotted as maps to correlate with topography and/or PFM images.[47]

As noted earlier, while here we discuss the specific case of the PFM, the same logic can be applied to any other imaging -spectroscopy methods pairs based on sequential spectral acquisitions, for example scanning tunneling microscopy and spectroscopy, optical imaging and nanoindentation, imaging and local chemical analysis via mass spectrometry, and many others. The key to the subsequent discussions are the timescales involved in the measurements, and the degree to which the measurement affects the material. These considerations affect the time budget and the feasibility of the experiment, respectively. For example, in nanoindentation measurements, mapping mechanical properties on a dense grid is impossible, since the material gets irreversibly deformed.

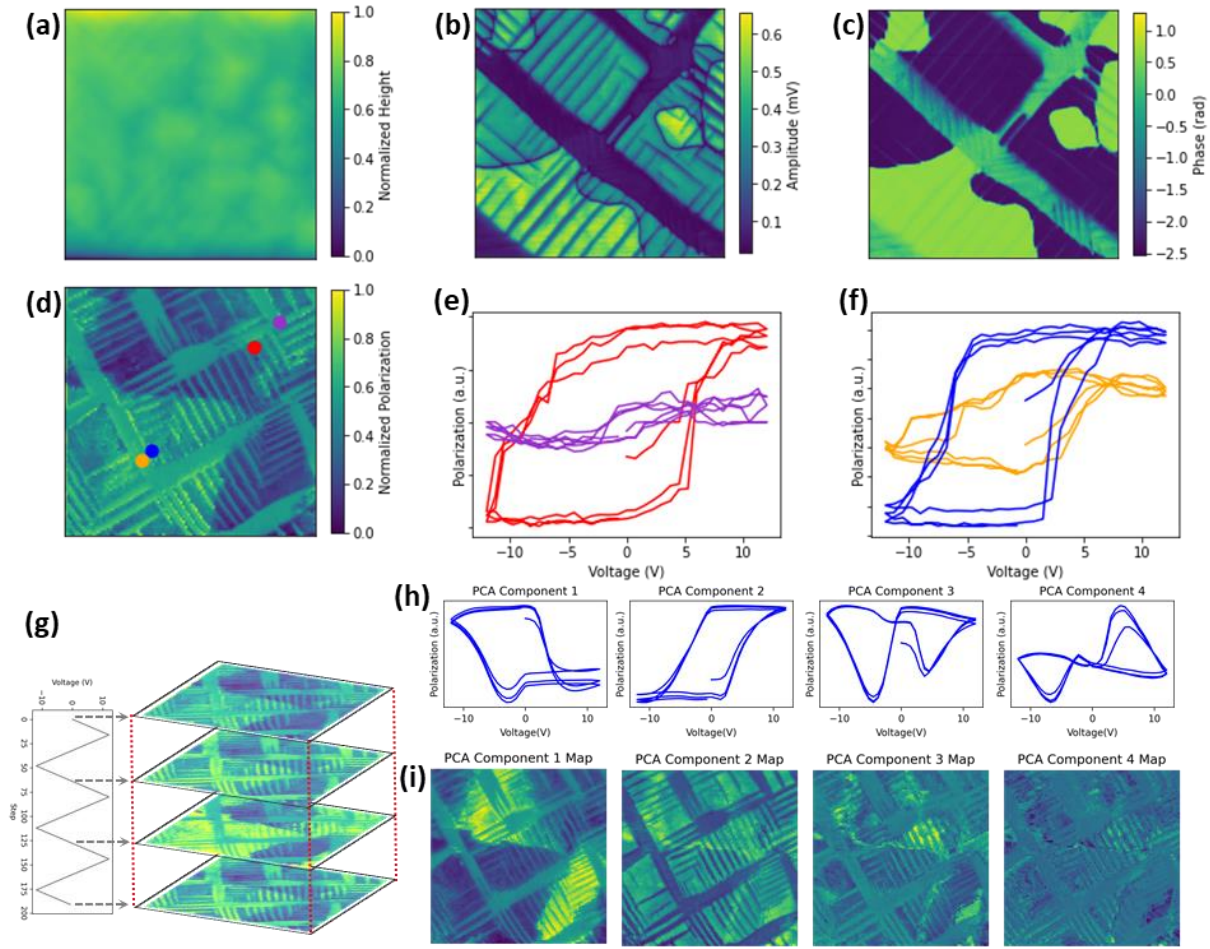

**Figure 2.** (a) Topographic and Piezoresponse force microscopy (b) amplitude and (c) phase images of the ferroelectric PbTiO$_3$ film. (d) Piezoresponse map of PFM switching spectroscopy data, (e), (f) hysteresis loops from selected locations in (d). (g) Piezoresponse force spectroscopy on the dense spatial grid of points gives rise to the 3D data set that can be visualized using dimensionality reduction methods, for example principal component analysis (PCA). (h) PCA components define the behaviors and (i) The PCA weights illustrate the spatial variability of specific behaviors. Adapted in part with permission from Ref [48], Copyright 2022 Springer Nature.

Consider an experimental scenario where we would like to explore specific functionalities of a material, for example polarization switching or mechanical properties, and determine the relationship between this property and the material's domain structure and/or topography. An example of PFM imaging and spectroscopy is illustrated in Figure 2.[48] Here, the surface topography image in Figure 2 (a) illustrates several topographical features associated both with the ferroelastic domain walls[49, 50] and the intrinsic material morphology. The corresponding PFM amplitude image shows the map of the electromechanical activity of the surface, with the in-plane domain being dark and out of plane domains being bright. The dark meandering lines on the image correspond to the ferroelectric walls between upward- and downward polarized out of plane domains.

Following the mapping of the topography and PFM images, the microscope can be configured to perform spectroscopic measurements. Originally, these measurements were implemented manually with the human-operator based selection of locations for spectroscopy.[51-53] In 2006, an approach for the spectroscopy grid measurements in PFM, or switching spectroscopy PFM, was introduced.[54, 55] In this, similar to force volume or CITS imaging, the individual spectra are collected over a rectangular grid of points and the resultant 3- or higher dimensional data set is analyzed by machine learning or physics-based methods.[56, 57] However, this approach can be limited by the measurement time. Secondly and even more importantly, the measurements can change the material (or tip), from subtle and often reversible changes of domain structure during polarization switching to irreversible damage using nanoindentation. While the changes in tip state are readily identifiable as contrast change and changes in material domain structure are easily detectable by standard PFM imaging after the hyperspectral data acquisition, the discovery of the structure-property relationships becomes impossible.

Hence, of interest is developing machine learning workflows that allow us to effectively sample the local materials behaviors based on topographic and PFM data, imitating the point-and-click approach of early days. One required component for such workflows will be the programming interfaces (engineering controls) that allow the updates to control signals such as tip position and initiation of the spectroscopic measurements. By now, several commercial microscopes allow for Python-based plug-ins as a part of the software architecture, and this trend is expected to continue . Alternatively, multiple work-around solutions based on partial control of microscope functions has been instantiated. The second component are post-acquisition analysis methods that can convert the measurements taken over the non-rectangular grids to the standard 2D images amenable to human perception. This can be accomplished via compressed sensing,[58, 59] Gaussian Processing (to be discussed below), or other interpolation methods. However, the third and key element is the selection of the measurement points, based on the information available before the experiment (i.e., topography and amplitude images, and perhaps data from similar samples), (potentially) decisions made upon observations during the experiment (e.g., human operator assessment of hysteresis loops), and the perceived reward of the experiment.

Here, we consider a typical exploratory strategy of a prototypical human operator. If interested in polarization switching at specific microstructural elements, the operator expects to obtain zero signal in the dark regions with purely in-plane polarization. However, given the uncertainty of the microscope performance, possible presence of the baseline signal due to the electrostatic interactions,[60-62] shear effects, and so forth, it is a good practice to acquire several measurements over the dark regions and make sure that responses are close to zero or consistent (and hence baseline can be subtracted). As a second step, the obvious next target for exploration are the regions with the stripe-like contrast corresponding to dense *a-c* domain regions that occupy most of the image. Here, the operator again explores the hysteresis loops as potentially related to pinning phenomena and ferroelastic wall dynamics. However, if the measurements yield consistent spectroscopic response over similar looking regions, the continuous experiments become redundant. With understanding of polarization switching behaviors in these regions achieved, the operator can proceed to explore other objects of interest, e.g., regions of ferroelectric domain walls with high curvature or crystallographically incompatible junctions between ferroelastic domains. For the former, we can expect to explore the role of flexoelectric coupling on wall properties while for the latter the local symmetry increases and transition to the paraelectric phases are possible. Note that while two first steps were driven by the motivation to characterize the instrument performance and explore statistically significant features within the image, the subsequent selection is predicated on the specific, and often highly specialized theories, within a specific scientific domain. The choice of these theories is informed by the prior knowledge of the operator. At the same time, implicit in this discussion is the concept of reward – which in this case is scientific discovery.

In this discussion, we illustrated that the experimental strategy in these cases is determined by several elements. It is informed by the prior knowledge of the human operator that interprets the image in terms of physical descriptors and builds a likely picture of responses. It is driven by the perceived reward of the experiment, whether pure exploration of possible functional behaviors, or interest towards specific object of interest. Curiously, rarely does the human operator consider the results of deep analysis of previous spectroscopy measurements during the experiment (beyond cursory visual examination) when selecting the location for the new ones. Hence, an interesting problem that is rarely explored is the search for microstructural elements that possess desired responses. With this introduction, we further proceed to explore possible strategies for automated experiment and introduce the sequence of Bayesian methods that incorporate data driven approaches and subsequently prior knowledge.

## 2. Experiment based on prior knowledge

It is instructive to discuss machine learning in an automated experiment, starting from examples of experimental workflows where objects of interest are known in advance. For example, defect identification in semiconductor wafers requires the full wafer surface to be imaged, requiring the scanning of multiple regions, stitching of the resultant images, and identifying the potential defect site(s). This problem is equivalent to the proverbial needle in a haystack and is solved via full grid search. It is important to note that if the defects are truly random, machine learning methods cannot offer any advantage in identifying them. Hence, in this case the automated experiment is equivalent to the grid search with a priori known target and no feedback, i.e., the experimental workflow does not change upon defect detection.

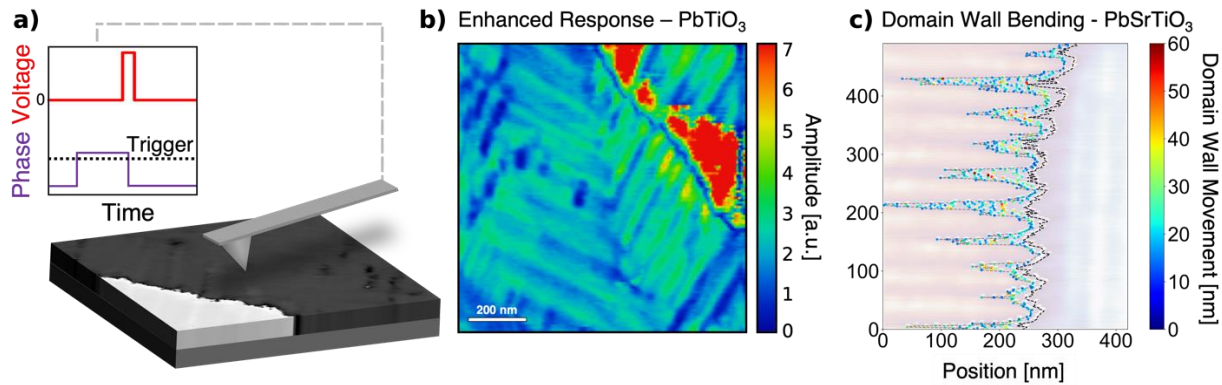

**Figure 3: Realtime feedback within a piezoresponse force microscopy**. a) Illustration of FerroBot triggering system utilizing a phase trigger and voltage pulse stimulus. b) Enhanced electromechanical response observed near pinned domain walls in PbTiO3 via FerroBot. c) Localized domain wall bending observed within PbSrTiO3 in-plane superstructure domains by deploying FerroBot. Adapted with permission from Ref [63], Copyright 2022 American Chemical Society.

Automated experiments with feedback can be realized without machine learning as well. The example of such an approach is the "FerroBot" (Figure 3) technique,[63] where the material state detected beneath the AFM tip is used as a real time feedback control for applying a stimulus, such as light, voltage, temperature, and pressure, in efforts to obtain localized control and modification of specific structures. Recently, we demonstrated enhanced electromechanical responses can be induced near pinned domain walls in PbTiO$_3$ via FerroBot, where the PFM phase is utilized as feedback for applying short (5 ms) voltage pulses in the areas of interest, i.e. at domain walls as shown in Figure 3b. Furthermore, we extended this technique within the previously developed framework to Pb$_{1-x}$Sr$_x$TiO$_3$ in-plane superstructures to gain fundamental insight into the dynamics of domain wall motion in such complex domain structures (Figure 3c).

This framework can be extended further with ML methods incorporated into the workflow. An example of this approach can be the recent work of applying a deep convolutional neural network-based YOLOv3 (You Only Look Once) system as an object detector to find the locations of molecules in AFM images, subsequently allowing acquiring high-resolution images of selected molecules.[34] Meanwhile, a Siamese neural network is applied to identify the same molecule in AFM images, keep tracking the imaged molecules and subsequently avoiding duplicated imaging of the same molecule.[34] As another example, a support vector machine (SVM), a supervised kernel-based method for non-linear classification, was applied to identified ferroelectric and electrochemical systems, making real time classification and performing additional probing at ferroelectric domain walls.[35] In scanning tunneling microscopy (STM), convolutional neural networks are employed to identified the tip condition, and upon the poor condition of tip, the program interfaces with microscope software to perform tip conditioning until the desired tip functionalization is achieved.[33, 36] The conditioning process can be optimized with reinforcement learning (RL) where anRL agent with sufficient training can automatically establish the rules of tip conditioning without human intervention.[37] Recently, we combined a deep convolutional neural network for segmentation of the domain structures with a rotationally invariant autoencoders to explore the domain dynamics in model polycrystalline ferroelectric materials and disentangle the factors affecting the pinning efficiency of ferroelectric domain walls. The ResHedNet identifies ferroelectric domain walls in PFM images..[64]

Here, it is important to highlight that the deep convolutional neural network (DCNN) used for the object recognition was trained prior to the experiment, and in this way represented prior knowledge in experiment planning. The state of the network did not change during the experiment and was not updated based on the experimental results. Similarly, the chosen actions upon the object detection were defined prior to the experiment. Hence, in this case the workflow utilizes ML component for detection of specific features but is not active learning.

The limitation of this approach is that pretrained neural networks tend to be susceptible to out-of-distribution effects. In this, the network trained for one specific imaging conditions will fail when the microscope imaging parameters change. This problem is now well recognized by the machine learning community in areas such as automated driving, medical imaging, and even dog vs. wolf recognition.[65-69] Several strategies for the minimization of the out of distribution drift have been explored based on the invariant risk minimization,[70] Barlow twin concepts,[71] and ensemble methods.[72, 73] For microscope imaging, possible strategies include the introduction of the translational and scale invariances into the DCNN architecture that account for possible microscope distortions. However, methods based on strong physical priors incorporated as inferential biases in the network (i.e., we expect atoms to be point-like objects in electron microscopy and domain walls to be continuous in PFM) have not been significantly explored. Still, these strategies allow only for partial improvement. Secondly, even if successful, these strategies are insufficient once we are interested in specific functional behaviors. For example, they are insufficient to explore what will be the behaviors of specific microstructural elements, e.g., hysteresis loops over specific domain configurations. However, these direct strategies are not suited to answer questions such as which domain structure gives rise to the maximal hysteresis loop opening or strongest imprint.

Hence, of interest are active learning methods where the knowledge of the system is updated during the experiment, and the experimental pathway is adjusted based on this new knowledge. From the ML perspective, this requires active learning methods, where the algorithm interacts with the data generation process rather than a static data set. Here, we discuss the basic principles and examples of Bayesian optimization based methods, one of the two major (along with the reinforcement learning) classes of active learning methods.

## 3. Basic principles of Gaussian processes and Bayesian optimization

To introduce the automated experiment based on the active learning methods, we first discuss the basic principles of the Gaussian Process (GP) and Bayesian optimization (BO). Currently, there are multiple excellent reviews and books introducing GP from the mathematical viewpoint.[74-77] However, very often these introductions are very mathematical in nature, and as the result, the basic principles of the BO, the intuition behind the BO, and perhaps even more importantly the limitation of existing methods, require considerable effort to understand. In this section, we introduce the GP first using a simple intuitive example, and then briefly illustrate the associated mathematical framework.

As a simple example, consider the (unknown) scalar observations $y$ on the 1D domain. These observations can represent the topography, current, electromechanical response along a single scan line in SPM, magnetization or polarization distribution along the 1D combinatorial spread library, or any other signal. The function $f(x)$ that generates these observations is unknown to the external observer, and the task is to learn it in the smallest number of steps (measurements). The measurements are assumed to be taken sequentially, with the location $x_n$ of a measurement at step $n$ chosen based on the results of measurements in the previous $n$ locations, $\{x_0, .. , x_{n-1}\}$.

Prior to the measurement, we do not know anything about the $f(x)$, which effectively means that the $y$ value is a random number and even the distribution from which these random numbers come is unknown. Clearly, this is not very helpful. The Gaussian Process framework makes two key assumptions about $f(x)$, namely it assumes that (a) the values of $f(x)$ belong to the same distribution and (b) the parameters of this distribution are correlated between adjacent locations. While the effects of (a) are not simple to illustrate and it rather postulates that we deal with the same type of process, (b) is very easy to illustrate as shown in Figure 4a where the bumps in $x$-range indicate that shorter distances have a stronger correlation. Now the function can still be anything, but there is obvious internal structure determined by the degree of correlations.

The natural question is how we would know the degree of correlation in advance. In the GP method, it is assumed that the correlations are controlled by a kernel of a certain functional form. The typical examples of the kernel functions are radial basis function (RBF) and Matern kernels:

$$k_{RBF}(x_i, x_j) = \sigma \exp(-d(x_i, x_j)^2/2l^2), \tag{Eq 1}$$

$$k_{Matern}(x_i, x_j) = \sigma \exp\left(-\sqrt{5} \times \frac{d(x_i,x_j)}{l}\right)\left(1 + \sqrt{5} \times \frac{d(x_i,x_j)}{l} + \frac{5}{3} \times \frac{d(x_i,x_j)^2}{l^2}\right) \tag{Eq 2}$$

where $d(\cdot, \cdot)$ is the Euclidean distance, $l$ is the kernel's lengthscale, and $\sigma$ is the output scale.

Note that all these kernel forms have parameters such as lengthscale $l$ in RBF and Matern kernels. In GP problems, the form of the kernel is generally chosen prior to the experiment and its parameters are learned from data during the experiment. At the same time, if the data was fully known, the kernel would be equivalent to the correlation function parametrized by the prescribed functional form.

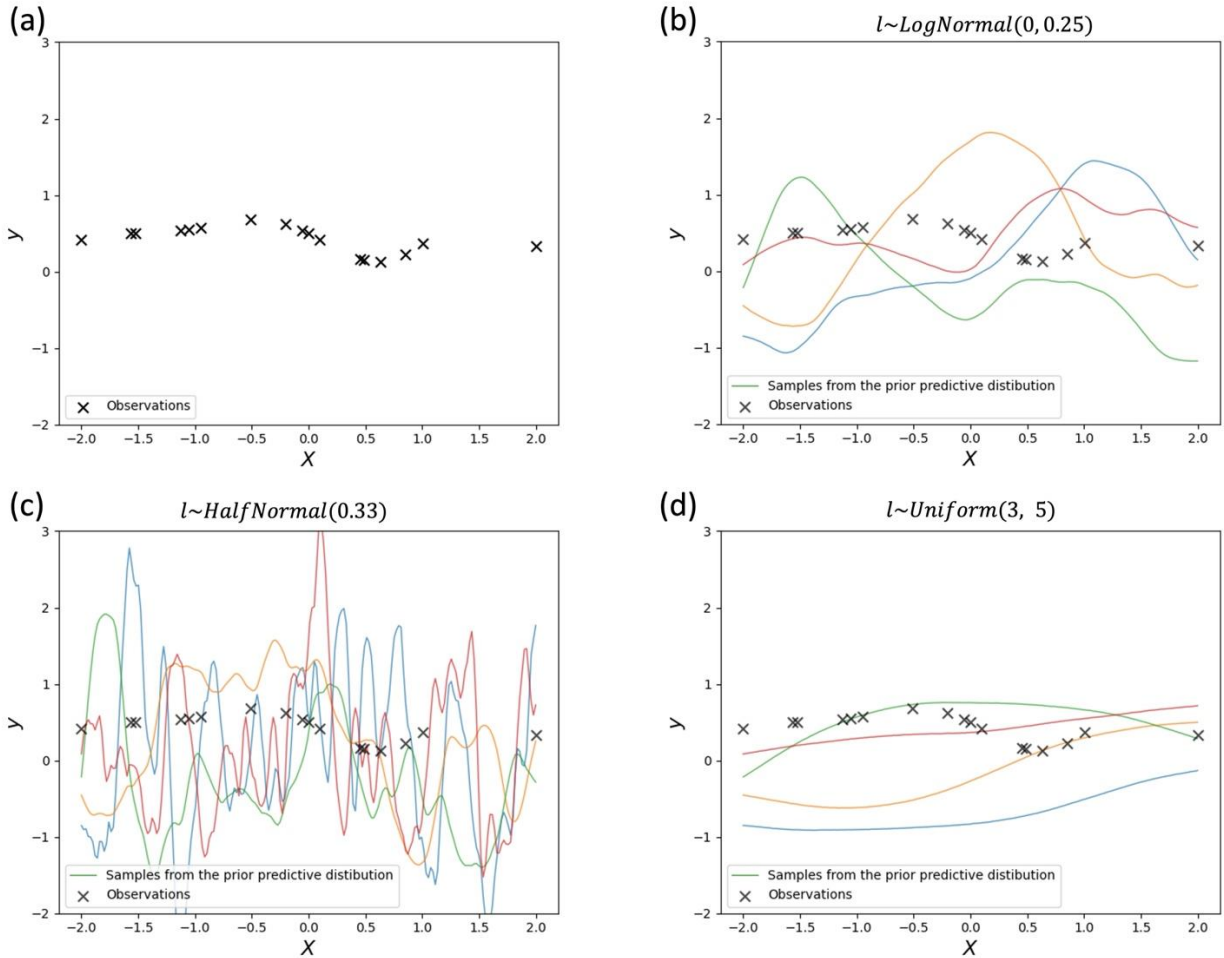

**Figure 4.** (a) Observations of a scalar function *y*. (b-d) GP predictions sampled from the multivariate normal *prior* (see Eq. 3-4), i.e., before conditioning on observations, for different prior distributions over the length scale parameter $l$ of the Matern kernel (see Eq. 2); assumes zero observational noise. The GP training is based on conditioning these prior predictive distributions on experimental data, i.e., only the traces that pass through the experimental points are chosen. Note the role of kernel length scale on the functions from the prior predictive distribution. In the GP, the kernel is learned from data.

Finally, the last step of the GP construction is the generation of posterior predictive distributions at the new data points. This process (assuming no noise) is illustrated in Figure 5. Here, the measurement at the location $x_i$ indicated by a cross mark that yields observation $y_i$ means that of all the possible functions only the ones for which $f(x_i) = y_i$ are real and the other functions are discarded. With this step, our function $f(x)$ is now defined at one location, but still can follow multiple forms away from it.

More formally, the GP can be written as a probabilistic model of the form,

$$f \sim MVNormal(m, K), \quad \text{(Eq 3a)}$$

$$y = f(x) + \epsilon \quad \text{(Eq 3b)}$$

Here *MVNormal* stands for multivariate normal distribution and *m* and *K* are prior mean and covariance functions. The former is usually set to 0 whereas for the latter, $K_{ij} = k(x_i, x_j)$ where *k* is a kernel computation function such as the ones in Eq 1 and Eq 2 for RBF and Matern kernels. The $\epsilon$ is a normally distributed observational noise. We then place priors on the parameters of the selected kernel. The priors on the kernel parameters are defined based on the prior knowledge of the scale of output values and the characteristic lengthscale of the phenomena of interest. In the absence of such knowledge, for Matern or RBF kernel, we can define a weakly-informative prior to ensure that both output scale and lengthscale are positive real numbers, e.g.

$$\sigma \sim LogNormal(0, 1) \quad \text{(Eq 4a)}$$

$$l \sim LogNormal(0, 1) \quad \text{(Eq 4b)}$$

In Figure 4b-d, we show samples from the prior predictive distribution defined by Eq 3, that is, before conditioning our GP model on observations (before "training" it). Note how the samples shape in Figure 4b-d vary depending on the chosen prior on the kernel length scale parameter (the output scale prior was the same in all cases).

The right choice of priors on kernel parameters may significantly speed up the model training, which can be critical in the live experiments. In our toy example, Figure 4b demonstrates a reasonable choice of the length scale prior, whereas the length scales in Figure 4c-d are either too small (Figure 4c) or too large (Figure 4d) for the observations at hand. The observational noise can be absorbed into the computation of the kernel function both at training and prediction stages.

Hence, the training process of GP model consists of inferring kernel parameters given the available observations. The inference can be done in a fully Bayesian way using the Hamiltonian Monte Carlo (HMC) sampling techniques, or it can be done using a maximum a posteriori (MAP) approximation. Multiple existing open-source software packages such as PyMC3,[76] Pyro,[78] and NumPyro[79] provide these inference tools. After inferring the GP model parameters, we can use it to obtain probabilistic predictions of function values on new inputs. Specifically, we sample from the multivariate normal posterior over the model outputs at the provided new inputs $X_*$ (usually, a denser grid of points):

$$f_* \sim MVNormal(\mu_\theta^{post}, \Sigma_\theta^{post}) \quad \text{(Eq 5a)}$$

$$\mu_\theta^{post} = m(X_*) + K(X_*, X|\theta)K(X, X|\theta)^{-1}(y - m(X)) \quad \text{(Eq 5b)}$$

$$\Sigma_\theta^{post} = K(X_*, X_*|\theta) - K(X_*, X|\theta)K(X, X|\theta)^{-1}K(X, X_*|\theta) \quad \text{(Eq 5c)}$$

where $\theta = \{\sigma, l\}$ are the inferred kernel parameters. Note that in the MAP approximation, we get a point estimate of kernel parameters and hence a single pair of $\mu_\theta^{post}$ and $\Sigma_\theta^{post}$. In the fully Bayesian mode, we get a pair of $\mu_{\theta^i}^{post}$ and $\Sigma_{\theta^i}^{post}$ for each posterior sample with kernel parameters $\theta^i$. Figure 5a shows the sampled predictions $f_*$ and posterior mean $\mu_\theta^{post}$ for our 1D example as semi-transparent red lines and solid blue curve, respectively. The regions where sampled predictions diverge, such as between $x$=1.0 and $x$=2.0, are referred to as regions with high uncertainty. The latter can be easily computed by taking a variance of the predicted values at each point.

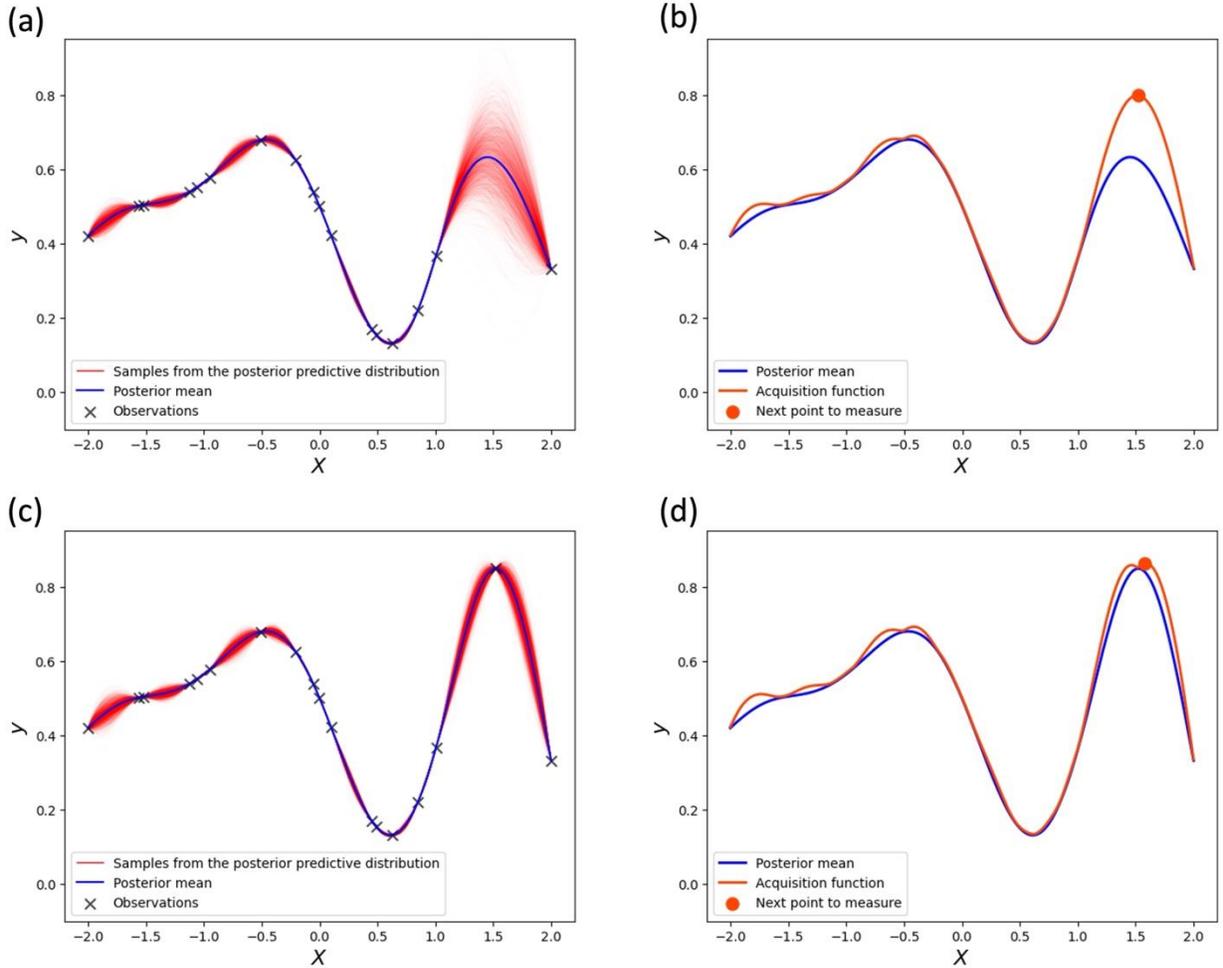

**Figure 5**. (a) GP predictions sampled from the multivariate normal *posterior* (see Eq. 5), i.e., after conditioning model on observations, and the GP posterior mean (Eq 5b). (b) The acquisition function derived from the posterior predictive distribution at a denser grid of points, according to Eq 7, with $c_2 = 1$ and $c_2 = 2$. The maximum of acquisition function determines the next point to measure. Note that practically, we compute acquisition function only for points that have not been measured yet. (c) GP prediction after measuring the suggested point and (d) the corresponding acquisition function values. Note that the *y*-axis is rescaled compared to Figure 4.

At this step, we incorporate the second key element of active learning. Specifically, the GP posterior predictive distribution is used to derive an acquisition function for selecting the next point to measure,

$$x_{next} = \arg\max_{\mathbf{x}} \frac{1}{M} \Sigma_{i=1}^{M} a\left(\mu_{\theta^i}^{post}, \Sigma_{\theta^i}^{post}\right) \quad \text{(Eq 6)}$$

where $a$ is a pre-selected acquisition function and $M$ is the number of posterior samples with kernel parameters. For the MAP approximation, $M = 1$. We can easily define our own simple acquisition function by taking a linear combination of posterior mean, $\mu_\theta^{post}$, and uncertainty, $\sqrt{\mathrm{diag}(\Sigma_\theta^{post})}$,

$$a = c_1 \mu_\theta^{post} + c_2 \sqrt{\mathrm{diag}(\Sigma_\theta^{post})} \quad \text{(Eq 7)}$$

where $c_1 = \{0, 1\}$ and $c_2 \geq 0$. The $\mathrm{diag}(\Sigma_\theta^{post})$ corresponds to the diagonal elements of the covariance matrix containing variances of the predicted variables whose square root is a standard deviation ('uncertainty'). We note that instead of $\mu_\theta^{post}$ and $\mathrm{diag}(\Sigma_\theta^{post})$ in Eq 7, one could in principle use the mean, $\bar{f}_*$, and variance, $\mathbb{V}[f_*]$, of predictions sampled from the multivariate normal posterior at each point.

When $c_1 = 0$ and $c_2 = 1$, we are in the **exploratory regime**. In this case, the algorithm seeks to minimize the total uncertainty about the system with the minimal number of steps. During the iterative process, the GP builds the reconstruction of the function, calculates the uncertainty, and suggest the next measurement point in which uncertainty is maximal. The process is iterated until uncertainty falls to a required level, or experimental budget (the number of points) is exhausted. When $c_1 = 1$ and $c_2 \geq 0$, we are in the optimization regime, that is, we want to discover regions where the value of the function is maximal – sometimes referred to as the **exploitation** problem. The pure, or greedy, exploitation problem ($c_1 = 1$, $c_2 = 0$) involves iterative measurements at the locations where the function is predicted to be maximal. The problem with this approach is that it gets trivially trapped into local maxima, without offering much advantages compared to simple gradient ascent. Hence, in practice, one uses the exploration-exploitation tradeoff, which is at the heart of the Bayesian optimization. For Eq 7, this corresponds to $c_1 = 1$ and $c_2 > 0$, and translates into the upper confidence bound (UCB) acquisition function.

We illustrate its application to our toy data in Figure 5b where the predicted function maximum is at $x \approx -0.5$. However, after considering the uncertainty in the prediction, the UCB acquisition function suggests exploring a point at $x \approx 1.5$ where potentially a true function maximum is located. We then perform a measurement in the suggested point and add the measured value to the initial training data. The updated training data is used to derive the new posterior predictive distribution (Figure 5c) and the corresponding acquisition function (Figure 5d). Here, after identifying what appears to be a true maximum, the algorithm suggests exploring points in its vicinity.

Another type of acquisition function frequently used in Bayesian optimization is the expected improvement (EI).[80] It estimates how much improvement one can get over the current 'best' point (for the maximization problems, it is the largest value observed so far). Unlike the UCB, the EI balances the exploration-exploitation tradeoff automatically. Other possible choices for the acquisition function are entropy search[81] and knowledge gradient,[82] although they can be too computationally expensive for online deployment on a microscope (i.e., potentially too high latency). A more in-depth technical description of the acquisition functions for Bayesian optimization can be found in the reviews by Frazier[83] and de Freitas.[84]

For the following discussion, we recap the major characteristics of the simple GP/BO algorithms, namely (a) we assume the existence of a non-observable ground truth function defined over a certain domain, (a) the measured data is assumed to be generated by the sampling from this function as an independent and identically distributed (i.i.d) process, (b) the spatial correlations are determined by a kernel of known functional form (c) the parameters of which is discovered from experiment and (d) the active learning process is driven by the balance of exploration and exploitation defined by the acquisition function.

## 4.  GP in experiment

With this introduction to GP, we now discuss its application in imaging and spectroscopy experiments. To define the automated experiment workflow, we cast the microscope operation to the Bayesian Optimization (BO) problem. For this, we define the parameter space, i.e., what are the experimental knobs available for the ML algorithm, and what is the function to be discovered/optimized. Note that the reward of the experiment is defined via the selection of the function to be optimized. For example, the experiment may seek to discover regions with maximal polarization and then the reward function is polarization itself, regions with minimal polarization, or polarization at some specific value. By the same token, the reward function can be more complex, e.g., regions with specific polarization pattern.

As an example, we can consider microscope tuning in intermittent contact mode. Here, we aim to optimize the microscope performance as a function of the imaging parameters, for example the setpoint value and driving amplitude. Based on the general physical considerations, we expect that for small setpoint values the area of the tip surface contact will be small and hence spatial resolution can be high. At the same time, long range forces will affect the measured signal, leading to smearing of the image. Comparatively, higher indentation forces result in stronger tip-surface interaction that now is dominated by local forces but lead to an increase of the contact area. Hence, in agreement with the experimental practices over 30 years of AFM history, we expect the optimal resolution to be achieved for certain intermediate values of these parameters. This microscope parameters define the parameter space of the BO problem. To avoid the risk to the equipment, we can introduce additional soft or hard restrictions on the extreme values.

The second component of BO is the function to be optimized. Here, it can be defined as resolution (which then needs to be determined form imaging or line scan data), noise, or frequency of the loss lines. With the function defined, GP will optimize the microscope resolution by dynamically adjusting the imaging parameters.

Another example of GP applications in microscopy is imaging or spectroscopic imaging. Traditionally, imaging in SPM is performed via raster scanning the region on the surface, giving rise to the rectangular array of sampling points. Hyperspectral imaging modes such as force volume measurements, CITS, or time- and voltage spectroscopies in PFM are performed on rectangular grids in a point-by-point mode. Hence, the natural question is whether more effective scan trajectories can be implemented for 2D scanning, and particularly, whether the locations for the spectroscopy measurements can be chosen more effectively. As such, we have recently demonstrated the use of coarse spiral scanning geometries during imaging to effectively increase the scan speed and reduce the tip-sample interaction (Figure 6). Specifically, we introduce the approach for fast scanning PFM based on the reconstruction of sparse spiral scans using two reconstruction techniques, compressive sensing and Gaussian processing, resulting in strong reconstructions with less than 6% error for GP reconstructions. Most notably, we achieve a high degree of image reconstruction accuracy while reducing the data collection by a factor of 5.8, ultimately enabling an increase in scan speed. Here, GP reconstructions were implemented for predefined sampling patterns. However, this technique can be extended to unknown or arbitrary scan patterns in conjunction with batch updates, i.e., the microscope collects a batch of points for determining updated locations for investigation.

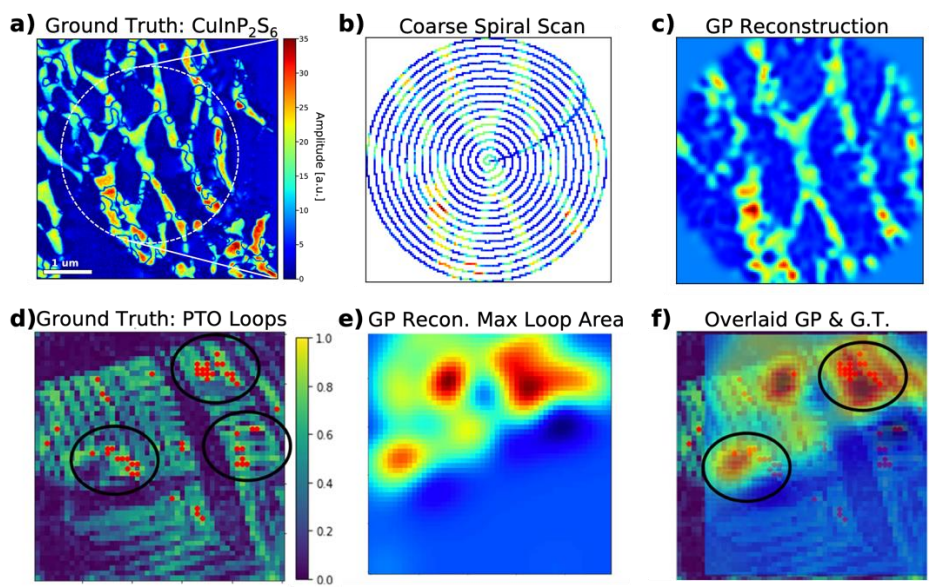

**Figure 6:** Gaussian process of scanning probe and electron microscopies. a) Ground truth PFM image of $CuInP_2S_6$, b) predefined coarse spiral scanning pattern with 16 spirals and the c) corresponding GP reconstruction. Adapted with permission from Ref [85], Copyright 2020 Wiley. d) Ground truth of high density piezoresponse spectroscopy grid illustrating spatial dependence of normalized hysteresis loop area. Note, red dots in panel (d) indicate locations of maximum normalized hysteresis loop opening (>0.8). e) GP prediction after 40 steps (400 pixels measured) of areas with maximum hysteresis loop opening. f) Panels (d) and (e) overlaid to illustrate accuracy of discovered large loop hysteresis area locations. Adapted in part with permission from Ref [44], Copyright 2021 American Chemical Society.

These approaches are illustrated in Figure 6.[44] Figure 6 (a-c) illustrate the GP reconstruction of the image based on a fixed sparse scanning pattern. Here, the scanning pattern is selected prior to the experiment, the signal is sampled across the corresponding pathway, and GP is used to reconstruct the image. The comparison between the ground truth (a) and reconstruction (c) illustrates the quality of this process – while the small details are lost, the large-scale features of the image are clearly visible. We note that in contrast to conventional computer vision approaches such as inpainting, the GP makes it easier to incorporate prior knowledge about the system and its prediction comes with quantified uncertainty for each predicted point, which can be used to decide where to measure next in the automated experiment.

Shown in the Figure 6 (d-f) are the results of the active GP learning of hysteresis loop area in the PFM spectroscopy measurements. Here, the hysteresis loop measurements were performed in specific locations, and the GP was used to identify the next batch of locations to perform the measurements. In this case, the GP was realized in the batch mode to allow for efficient use of the computational resources. Note that the comparison of the reconstructed image with the ground truth image (i.e., acquired over the dense grid of points) suggest that the active experiment results are fairly poor – while the large-scale features are reproduced, the fine contrast details associated with specific domain walls are not captured in GP reconstruction.

While continuing the GP driven active learning experiment to many experimental samples can yield better results, there are several significant limitations. First and foremost, GP methods are computationally intensive and in the context of experiment of Figure 6 (d-f) required the use of a GPU server. Secondly, in SPM the scanning process is associated with the nonlinearities and drifts, meaning that exploring the sample surface in arbitrary sequence introduces a certain level of positional disorder. Furthermore, the use of batch updates further necessitates the selection of pathfinder function, i.e., defining the sampling sequence under the constraints of microscope operation, representing a known example of NP hard problem that we never come back exactly where we were. Finally, GP will still be expected to require a considerable number of sampling points, and although it can reduce experimental budget to 10-30% of grid measurements, the tradeoff is that it renders the experimental procedure considerably more complex.[86]

Overall, the results shown above yield several conclusions. For scanning with non-rectangular trajectories with fixed sampling points, the GP method allows for image reconstruction and offers less then order of magnitude improvement in sampling. However, for the active experiment the gain in terms of minimizing the number of sampling point is comparable, whereas the complexity of the experiment both from the instrumental and computational perspective is considerable. Overall, the use of simple GP for active learning in imaging clearly did not bring advantage sufficient to make it a part of the experiment.

The origins of this behavior can be traced directly to the basic postulated of GP as discussed in section 3. In the examples above, we aimed to explore the black box function using either predefined measurement sequence, or active learning experiment. The GP learns to interpolate the function of interest via learning the kernel function, which is effectively a parametrized form of the correlation function. The image as shown in Fig. 6 contains multiple details at all length scales which cannot be defined via a single correlation function. Hence, the rough details are reconstructed, but finer details associated with different length scales remain invisible. More importantly, since simple GP uses a single kernel, discovering the large-scale details will reject the finer details, effectively acting as a low-pass filter.

One of the ways to improve the GP/BO is to introduce more complex kernel functions allowing for the different functional forms in the image or choose a mixture of kernels so that the correct kernel and its parameters are learnt during the experiment. In this case, the kernel function will contain the information about materials. For example, spectral kernels can be well-suited for images with atomic periodicity that can naturally be represented as superposition of the Fourier components. However, this direction rapidly becomes computationally intractable. From the experimental viewpoint, the choice of kernel then starts to be predicted on the expectations on system behavior, i.e., rely on experimentalist biases.

Hence, the simple GP/BO method while being very useful for microscope tuning and fixed-point reconstructions, are insufficient for more complex active learning workflows. The reason for this is that they are purely data driven strategies that impose certain constraints on the function to be discovered, and these constraints are purely correlative in nature. Comparatively in most applications, we have access to other data or some information on physical laws acting within the system. Examples of the BO strategies that use these methods are discussed in subsequent sections.

## 5. Deep kernel learning for automated discovery

In microscopy (or any other) experiment, we almost never consider the problem of black box exploration, thus immediately going beyond the examples in Section 4. When selecting material, we typically have anticipation of what type of behaviors it is likely to manifest and have specific research goals in mind. These considerations define our prior knowledge before the experiment and reward that drives the experiment. On a more local level (once material and reward are selected), the same principle applies. Typically, we utilize successive measurement to obtain progressively deeper insight into behaviors of interest by choosing next regions for exploration. For example, in SPM experiments, optical microscopy is often used to identify the regions of interest for detailed studies. Similarly, once the topographic/functional maps of the surface are obtained, they can be used to identify the regions of interest to perform spectroscopic measurements. In the hysteresis loop measurements experiment exemplified in Section 2, the observed domain structure was used to select locations for spectroscopic measurements.

In the following discussion, we disambiguate two very important scenarios for automated experiments. In the first scenario, we know in advance what objects of interest in the topographic or functional image we are interested in. For example, we may choose to explore the polarization switching at the domain walls as a function of local surface curvature, or any descriptor that can be derived from observations. This corresponds to the case discussed in Section 2, and we refer to it as direct automated experiment problem.

The second case is the case of automated discovery (or inverse automated experiment problem), where we are interested in certain spectroscopic responses, and aim to discover the microstructural elements at which they manifest, see Figure 1. This problem can be cast as a pure exploratory discovery experiment, in which we aim to maximize the variability of observed responses. Alternatively, the experiment can be developed so that the algorithm targets the discovery of regions on the sample surface where the response satisfies certain criteria. These experimental workflows can be realized via deep kernel learning (DKL). Below, we briefly summarize the DKL principles and illustrate its application for ferroelectric materials discovery.

DKL can be represented as a combination of deep neural networks and Gaussian Process. Here, the neural network is used to discover the low-dimensional representation of data, commonly referred to as embedding vectors or latent variables. The Gaussian Process operates over the discovered low dimensional space of the latent variables. Formally, the deep kernel is defined as

$$k_{DKL}(x_i, x_j|w, \theta) = k_{base}(g(x_i|w), g(x_j|w)|\theta) \qquad (Eq\ 8)$$

where $g$ is a neural network with weights $w$ and $k_{base}$ is a standard GP kernel such as Matern or RBF. Hence, the deep kernel can be used as a drop-in replacement of the standard GP kernel in Eq. 3-5. We also note that an important difference between DKL and the GP applied to the PCA, or NMF-reduced image data is that in DKL, the parameters of the neural network and GP are inferred jointly, in an end-to-end fashion, and therefore, the 'reduction' of high-dimensional image data is conditioned on the observed physical functionality.

The DKL-driven experiment proceeds as follows. In the first stage of experiment, the microscope collects the surface topography (or functional) 2D image over the relatively large field of view. This is considered to be a relatively 'fast' measurement. With this information in hand, the algorithm represents the image as a collection of patches centered at individual pixels (Fig. 7a). These patches form a spatial structure descriptor. Based on the nature of the measurement, the patch can be selected from one information channel, or represent several channels. For example, in band excitation Piezoresponse Force Microscopy the patch can be NxNx1 PFM amplitude signal, or NxNx3 amplitude, phase, and resonant frequency stack. For MxM image, this procedure yields $\sim(M-N)^2$ patches. Note that while more complex spatial descriptors can be used (e.g., via the positional embeddings, etc.), these approaches have not yet been explored for physical data analysis.

Next, we select a *scalarizer* function that will convert a measured spectrum into a scalar physical descriptor. The acquisition of such spectra is considered to be a 'slow' measurement. The scalarizer defines the measure of physical interest to the response and acts as a reward towards the specific behavior in the experiment. The scalarizer can be created in almost unlimited number of ways. For example, it can be based on the gross characteristic of the spectrum, such as area under curve, integrated intensity within certain energy range, area or width of hysteresis loop (Fig 7b). It can incorporate the physical model and physics-based analysis, for example converting predicted spectrum to specific materials parameters. It can be based on hybrid criteria defined via combinations of functional fits, decision trees, etc. Finally, it can be crowd sourced – if the neural network has been trained by human labelled data, or other form of expert system can be used. Ultimately however, the scalarizer should define the measure of physical interest much like the human operator would do.

To initialize the DKL algorithm, a few spectrums are collected at randomly selected pixels. With this, the DKL algorithm is trained using patches centered in those pixels as inputs and the scalarized spectrums as targets. This allows DKL to learn a probabilistic relationship between these available input patches representing structural features and physical behavior of interest. The trained DKL model is then used to predict the property of interest in for the remaining image patches (for which there are no measured spectrums yet). We note that the DKL model can be also used in the so-called multi-output regime to try reconstructing the entire spectrums over the field of view, but the computational cost is usually prohibitive for the implementations in real time. Because the DKL prediction is probabilistic, it can be used to derive the acquisition function (see Eq. 6) for selecting the next measurement location (image patch). Once the next measurement is performed, the corresponding image patch and the scalarized spectrum are added to the training data and the process is repeated iteratively until experimental budget is exhausted or the required predictability is achieved.

An example of DKL-based automated experiment in scanning probe microscopy is shown in Fig. 7(c-e). Here, the structure-property relationship of a $PbTiO_3$ ferroelectric thin film is explored with DKL driven band excitation piezoresponse spectroscopy (BEPS) measurement. The piezoresponse force microscopy amplitude image showing ferroelectric and ferroelastic domain structures is used to created structural descriptor here, and the scalarizer physical descriptor is the loop area of piezoresponse vs. voltage hysteresis loop. In this case, the DKL actively learns the relationship between local domain structure and corresponding hysteresis loop area. The results in Figure 7d and 7e show the DKL exploration points based on on-field hysteresis loop area and off-field hysteresis loop area, respectively, which indicate that DKL discovered the object of interest is ferroelectric and ferroelastic domain walls. Noteworthily, DKL model does not contain physical knowledge and there is no prior information about the importance of domain walls in a ferroelectric material; nonetheless, DKL discovered that the object of interest in this material is domain wall during the measurement. In addition, the different exploration points under on- and off-field conditions—exploration points concentrate around 180º ferroelectric domain walls under on-field condition (Figure 7d) and around non-180º domain walls under off-field conditions (Figure 7e) under off-field condition—suggest that the structure–hysteresis relationship varies under different circumstances (on- or off-field).

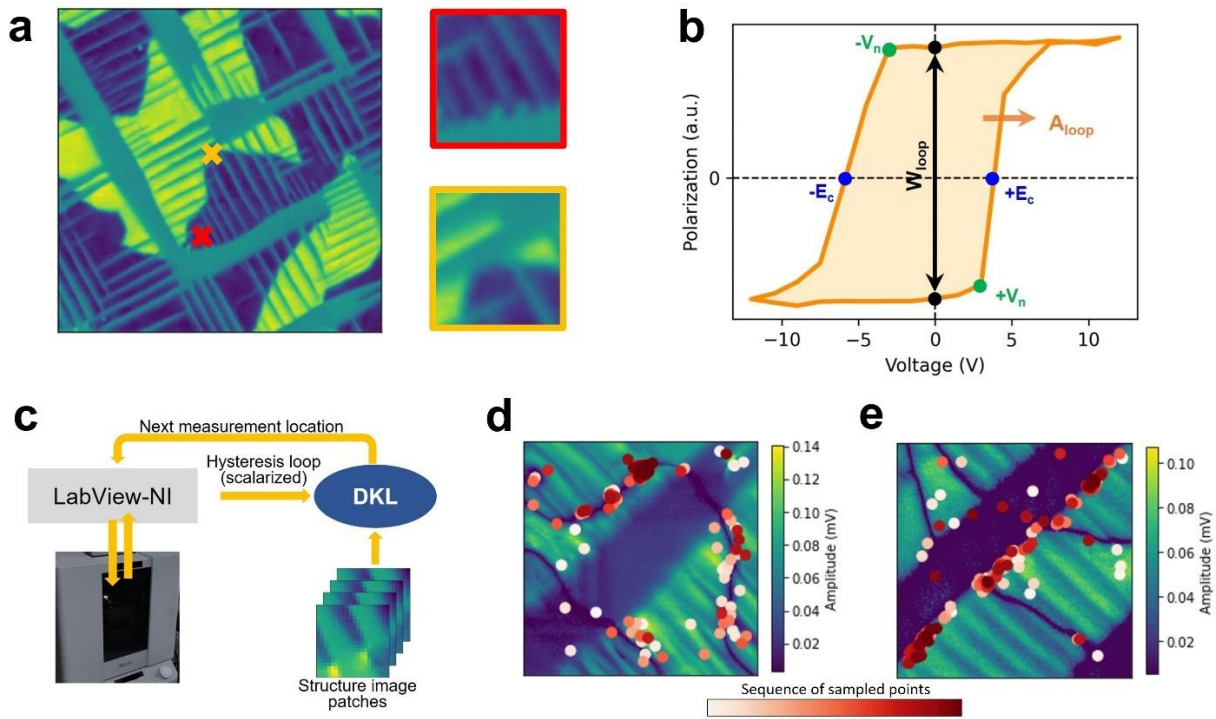

**Figure 7**. DKL-based automated experiment in scanning probe microscopy. (a) An experimental ('structural') image over a relatively large field of view is split it into a set of patches extracted at each pixel. (b) The scalarizer function is created to convert a spectrum (measured in each patch) into a scalar physical descriptor associated with a physical behavior of interest. (c) At each step of the automated experiment, the DKL model learns a relationship between image patches for which the hysteresis loops are already measured and the corresponding property of interest (extracted from those loops via a scalarizer) and makes a probabilistic prediction about the next measurement location. (c) schematic showing implement of DKL in operating SPM. (d, e) Example of the DKL-sampled points during the automated experiment in the on- (c) and off- (e) field regimes overlaid with a BEPFM amplitude image. Adapted in part with permission from Ref [48], Copyright 2022 Springer Nature.

In cases where multiple image channels are available (e.g., topography, phase, etc.), the DKL can be used to identify channel with the best predictive capacity of the physical property of interest based on the predictive uncertainty. Generally, the channel with better predictive capacity leads to the faster decay of the overall posterior variance, which allows discarding some of the redundant channels in the automated experiment.

**6.    Bayesian Inference for structured probabilistic models**

With the basis data driven GP and its extension to the DKL introduced, we discuss the second aspect of the Bayesian methods, namely Bayesian inference. One of the common tasks in data analysis is fitting the data to a specific model. In band excitation PFM, we aim to reconstruct the parameters of the simple harmonic oscillator from the amplitude-frequency curve. In AFM force-distance measurement, the indentation curve is used to calculate the elastic properties of materials, whereas force in the non-contact regime are used to estimate the long range electrostatic or magnetic interactions.

The classical and well familiar approach for analysis of such data when the model, meaning the analytical or numerical relationship linking the parameters of interest and measured data is known, is the least squares (LS) approach. In this approach, we chose the function *f(x)* which can be evaluated at points $x_i$ to yield function values $f_i$, and define the mean squared error (MSE) as

$$MSE = \sum_i [y_i - f(x_i, a_1, \ldots a_n)]^2 \tag{Eq 9}$$

During the fit, we aim to minimize the MSE by varying the parameters of the model $a_1, \ldots a_n$. This utilizes an optimization algorithm such as the Levenberg–Marquardt[87] algorithm, which approximates the changes to function values by a first order (linear) change based on the computed gradient of the function with respect to each fitting parameter. If the parameter values are chosen close to their true values, this method converges to the global minima; but if multiple minima are present and the initial priors input to the algorithm are not close to their ideal values, local minima will instead be found. Once the optimization is complete, the parameter values are assumed to represent the values that "best" fit the physical equation provided.

The delineated approach has several weaknesses. Practically, choosing the parameters well outside of their true values often leads to the spurious fits. Secondly, the functional form of the behaviors of interest is not always exactly known, and it is common to compare fits by several models and choose the one that offers least error. However, on an even more fundamental level, the least square fitting method does not allow one to express confidence of possible parameter values before the fit is undertaken. In other words, the parameter can only be taken as either free or fixed. In some implementations, it is possible to add rigid constraints (e.g., non-negativity). After the fit, parameters can be compared to physical expectations, and based on these, the model can be accepted or rejected. For example, cantilever resonance peaks should have positive quality factor and amplitude, for fitting optical dielectric constants, we expect positive values, etc. However, this physical prior knowledge is used during the interpretation stage, but not normally factored into the analysis.

An elegant and universal way to incorporate prior knowledge into the fit is via the use of Bayesian Inference (BI) methods. In BI, we assume that we know the relationship of the model connecting the measured signal to system parameters. However, the parameters are now given in the form of their prior distributions. If a certain parameter is very well known, the corresponding distribution can be taken to be very narrow. If the parameter is poorly known, the distribution can be a broad Gaussian, Semi-Gaussian (if e.g., positively defined), or uniform (if constrained within certain interval). Importantly, the prior distributions can be estimated much more precisely from previous measurements, literature data, etc. Furthermore, they can be also multimodal. For example, for ferroelectric material the polarization can be +P and -P, and the corresponding prior distribution is bimodal.

The combination of the model and prior distributions on the parameter values is defined even before the experiment and defines the prior knowledge about the system. The experimental data provides new data on the system behavior. The essence of the BI method is that given the new data, the priors are updated to yield posterior distributions, which are narrower than the prior. This narrowing reflects the increased knowledge about the system derived as the result of the experiment.

Note that in the classical statistical community, this was traditionally perceived as the limitation of BI, i.e., that it required us to put prior distributions on variables. This consideration was traditionally perceived as a weakness compared to the frequentist (i.e., only experiment-based measures). However, in the physical sciences this knowledge is inherent, and hence BI is ideally matched to applications in the physics domain. Perhaps even more importantly, the in-depth comparison of the Bayesian and frequentist methods suggest that many of the common statistical criteria (*p*-values, etc.) implicitly rely on specific prior hypotheses, as analyzed in detail by Kruschke.[88] However, these frequentist methods lack flexibility to express the complex prior structure.

In both frequentist and Bayesian approaches, a common challenge is that there is not one model but multiple models that can be used to describe any given dataset. In such circumstances, it becomes important then to determine which model is most likely given the data. Of course, simply computing and comparing the loss via equation (1) amongst different models is not particularly helpful, because it will favor more flexible models over those with less degrees of freedom. As such, one needs a metric that can compare different models with respect to the number of degrees of freedom and penalize those extra degrees of freedom for each additional reduction in loss they provide.

There are a variety of approaches, such as the Akaike information criterion (AIC), but a recently developed approach by Watanabe, termed the widely applicable information criterion (WAIC)[89], extends this thinking to a more strictly Bayesian setting. Here, assume we have several models $M_k = M_1, \ldots, M_N$. Each model has $i$ parameters $\theta_i^k$, each parameter having a prior distribution that can be informative or non-informative. We then compute the posteriors of the parameters based on the datasets to obtain posteriors over the parameters $\theta_i^k$. This step is the most computationally expensive, and usually undertaken with some form of sampling, such as Markov-chain Monte-Carlo (MCMC), Hamiltonian Monte-Carlo, No U-Turn Sampling, or other approaches. With these posteriors, we can calculate the log likelihood (LL), which is simply the probability of datapoint $y_i$ given some model parameters $\theta$, and for the Bayesian case, can be written as[90]

$$LL = \sum_{i=1}^{n} \log\left(\frac{1}{S}\sum_{s=1}^{S} p(y_i|\theta^S)\right) \quad \text{(Eq 10)}$$

Where there are $S$ simulation draws and $n$ datapoints in total. Given that we already have the posteriors for the parameters, we can use those in the simulation draws to easily compute the LL. We then compute the penalty term for the degrees of freedom in the model (effective number of parameters), which we denote as $p_{WAIC}$, and is calculated by

$$p_{WAIC} = \sum_{i=1}^{n} \text{var}_{post} \log(p(y_i|\theta)) \quad \text{(Eq 11)}$$

Where var is the variance and can be calculated as $\text{var} = V_s^S a_s = \frac{1}{S-1}\sum_{s=1}^{S}(a_s - \bar{a})^2$. This is used precisely because the effective number of parameters strongly depends on the variance of the parameters. Then the WAIC is calculated as WAIC = LL – $p_{WAIC}$. The probability of the models $p(M_i)$ can then be calculated via an approach defined in ref[91] which uses an AIC-type weighting approach with Pseudo-Bayesian model averaging. The advantage of the fully Bayesian model selection approach here is twofold. One, as discussed earlier, it enables us to incorporate priors on different model parameters, and therefore better inform the probabilities of the models, and second, it is more applicable for hierarchical models than the traditional point estimator approaches. In contrast to the traditional frequentist approach, the other relevant observation here is that the uncertainty of individual parameters within the model directly affects the model probability.

To show its utility, a subset of authors has performed such Bayesian model selection in the case of both scanning probe microscopy (SPM) as well as electron microscopy. In SPM, we utilized this approach for determining whether a linear or nonlinear oscillator would better describe observed cantilever deflections on a thin film $PbTiO_3$ sample,[46] with Landau modeling suggesting that ferroelectric behavior could result in the observed mechanical nonlinearities. In Figure 8(a), simulated data is used to show how the traditional AIC scores vary as a function of cantilever nonlinearity and noise level, as compared with the smoother changes in the WAIC approach in Figure 8(b). An example curve from real data is shown in Figure 8(c), with the data as red circles, the (linear) simple harmonic oscillator fit in blue, and the (nonlinear) Duffing oscillator in black. By eye, it appears both models are very well suited to this data; however, WAIC scores consistently favor the Duffing oscillator model.

The second example concerned atomically resolved images of domain walls in a $BiFeO_3$ thin film sample[92] with three distinct models for ferroelectric polarization being explored via Bayesian inference and model selection. This is illustrated in Figure 8(d-i), with the atomically resolved image of the 180° domain wall shown in Figure 8(d), the (perpendicular) component the polarization shown in Figure 8(e), and the wall profile shown in Figure 8(f). Running Bayesian inference enables posteriors to be retrieved based on this data; the posteriors are plotted in Figure 8(g-i) and show that although some parameters are well defined with narrow, sharp peaks, such as the spontaneous polarizations $P_s$, others are more uncertain, such as the correlation length $L_c$. Practically, extending the use of these model selection approaches will require integrated computation-experiment workflows given the heavier computational demands of Bayesian inference.

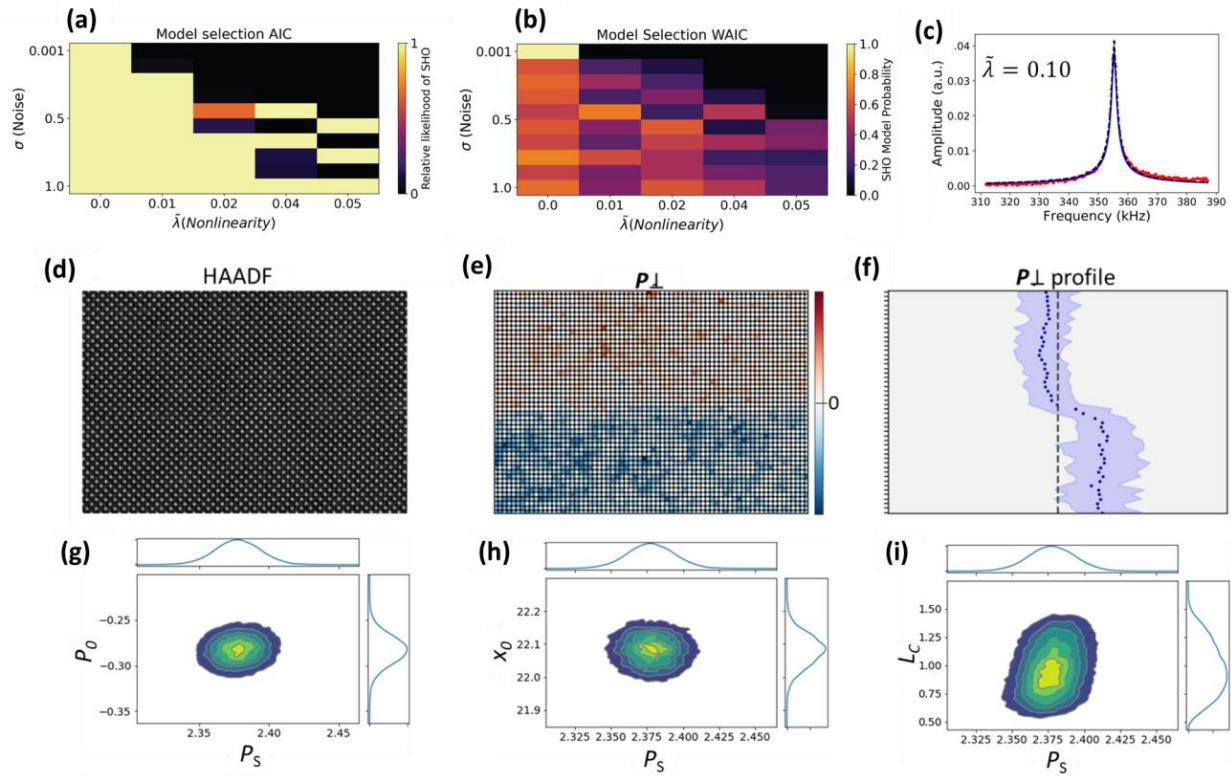

**Figure 8.** Bayesian inference and model selection in microscopy. Model selection in scanning probe microscopy is shown in (a-c). Here, we compare the traditional Akaike information criterion approach (a) with the more Bayesian approach (WAIC), in (b), for simulated data. The data is that of a nonlinear oscillator with varying levels of noise (vertical axis) and nonlinearity of the oscillation (horizontal axis), and the model selection is between that of a simple harmonic oscillator model and a Duffing (nonlinear) oscillator model. Color represents probability of the linear model. (c) Example spectrum, with the raw data in red, the simple harmonic oscillator prediction in blue and the Duffing oscillator prediction in black. An example of Bayesian inference for electron microscopy data is shown in (d-i) with the high angle annular dark field image of a 180° domain wall in BiFeO3 shown in (d), the associated (perpendicular component) polarization map in (e) and the wall profile extracted from this map in (f). Blue dots indicate the mean data and the color indicates 90% extend of the data. Posteriors after Bayesian inference are plotted for a particular choice of model in (g-i). these posteriors show considerable uncertainty around the parameter $L_c$,

the correlation length, whereas the spontaneous polarization appears known to significantly more confidence. Adapted with permission under a Creative Commons Attribution 4.0 International License (CC BY 4.0) from Ref [92].

## 7. Structured Gaussian Processes and Hypothesis learning

The Bayesian optimization approaches described above were either purely data driven, i.e., relying on the data obtained during the same experiment, or were using the proxy data to learn structure-property relationships. However, almost universal in many areas of physics, chemistry, and materials science is the existence of certain physical laws that describe the behaviors of interest across the corresponding parameter space. In rare cases such as Newtonian mechanics or general relativity, these laws are precisely defined. In many other cases, there are common behaviors including open parameters, for example Arrhenius laws for the reaction kinetics. Finally, in many cases the laws are largely phenomenological in nature. However, common to all these scenarios is that certain definitive statements about the behavior of the system of interest can be made at a different level of certainty. For example, when exploring temperature dependence of heat capacity in unknown material, some general assumptions can be made regarding the possible values of heat capacity (non-negative, limiting values) and its behavior close to 0K. If it is known (for example, from structural data) that the material has a phase transition at certain temperature, then the likely behavior of heat capacity includes a peak, with the exact shape determined by material purity and order of the phase transition. Note that while general, these statements provide a very well-defined structure for the possible behavior of the system, - more constrictive and detailed than simple GP assumption of specific correlations. Comparatively if we use purely GP methods for exploring heat capacity, none of these behaviors are considered. The simple GP/BO methods are purely data driven.

However, the nature of Bayesian methods is such that this information can be easily factored in. In the mathematical formalism of Section 3, the GP process was set as the function with zero mean and correlations defined by the kernel (see Eq 3). However, the prior mean function can be non-zero and in fact represent the physical behaviors of interest. Interestingly, the vast majority of the reviews and books on Bayesian methods focus the discussion on kernel structure analysis, while deferring the mean function to the domain areas.[74] However, it is this flexibility of the GP that allows physical knowledge incorporated in the model.

To incorporate our prior knowledge into GP/BO, we substitute a prior mean function $m$ in Eq 3 with a probabilistic model of expected system behavior whose parameters are inferred together with the kernel parameters. Then, the posterior mean in Eq 5 becomes

$$\mu_{\theta^i \phi^i}^{\text{post}} = m(X_*|\phi^i) + K(X_*, X|\theta^i)K(X, X|\theta^i)^{-1}(y - m(X|\phi^i)) \quad \text{(Eq 12)}$$

with the rest of the GP equations unchanged. The $\phi^i$ is a single posterior sample with learned parameters of the probabilistic model *m*. Hence, the uncertainty about model parameters is now automatically considered during the active learning and Bayesian optimization. This results in a semi-parametric form of the Bayesian optimization and active learning algorithms allowing for the 'physics-informed' decisions about which points to measure next. The incorporated probabilistic model doesn't have to be precise, but it should capture general or partial trend in data. Note that in a sense structured GP can be interpreted as a "hybrid" of the Bayesian Inference in section 5 and the simple GP process in section 4, inheriting the strength of both approaches and avoiding their shortcomings as will be discussed below.

In Figure 9 we showed an example of modelling a phase transition behavior using a vanilla GP, the structured GP with correct model of physical behavior, and the structured GP with a partially correct model of physical behavior. Clearly, injecting even a partial knowledge about the system significantly improves the GP prediction.

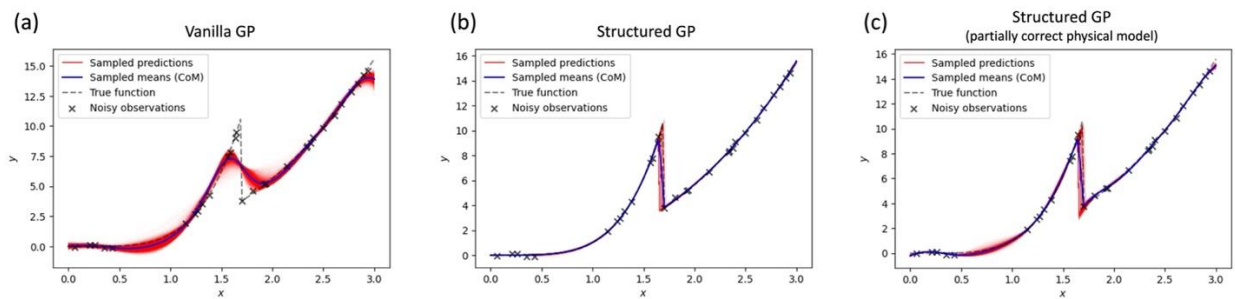

**Figure 9.** Predictions of vanilla GP (a), structured GP with a correct model of physical behavior (b), and structural GP with a partially correct model of physical behavior (c) to a dataset with a phase-transition-like behavior. In (b), we used a piecewise function with power law behavior before and after the phase transition, with normal priors over the power law coefficients and uniform prior on the position of the transition. In (c), we used a piecewise function with linear behavior before and after the phase transition point, with normal priors over the slopes, and uniform prior over the position of the transition. Note that even partial knowledge of salient features of true function significantly accelerates the discovery process.

The structured GP can be used for hypothesis learning in automated experiments. The hypothesis learning is based on the idea that in active learning, the correct model of the system's behavior leads to the faster decrease in the overall uncertainty about the system under study. Specifically, in the hypothesis learning setup, several competing (probabilistic) models of expected system's behavior are wrapped into structured GPs, and a basic reinforcement learning policy (e.g., epsilon-greedy) is used to select a correct one based on the collected 'rewards'.

Recently, we demonstrated an application of the hypothesis learning to the ferroelectric domain switching in PFM. In these experiments, the material is initially switched to a single domain state via scanning with biased tip. Subsequently, the bias pulse of a certain length and amplitude are applied to a selected location and resulting polarization pattern is imaged. Depending on the pulse parameters, the oppositely polarized domain can form. The resultant image can be analyzed to yield the domain size (for round domains). The dependence of the domain size as a function of pulse amplitude and length constitute the domain growth law.

Historically, the bias- and amplitude dependence of domain size were explored via direct grid measurements, where a series of pulses with varying length and amplitude were applied to the surface, domain sizes were analyzed, and subsequent dependence of $R(V,\tau)$ was analyzed.[93, 94] At the same time, a number of models for $R(V,\tau)$ corresponding to different limiting cases of switching were proposed. Here, Molotskii,[95, 96] Kalinin,[97] and Morozovska[98, 99] developed the thermodynamic theory of domain switching for different boundary conditions on the surface. In this thermodynamic limit, the size of the domain is independent on time and is determined solely by materials and surface properties and bias magnitude. At the same time, the thermodynamic theory has elucidated the critical role that surface screening plays in domain size evolution. This observation combined with the known presence of slow charge migration on oxide surfaces in ambient conditions[100-102] suggest that domain growth kinetics can be controlled by the diffusion of the screening charges, as was convincingly demonstrated by Yudin.[103] In this approximation, domain size will be determined by the interplay of electrochemical charge generation at the tip-surface junction and lateral diffusion and electromigration of screening charge in the tip field. Finally, the third possibility is the kinetic model in which the domain growth kinetics is controlled by the pinning of the domain wall in the random impurity potential, as explored in depth by Paruch.[94, 104, 105] Correspondingly, these models yield the limiting cases for the domain growth dynamics in automated PFM experiment. For all scenarios, the alternative simplified forms of the domain growth kinetics are available.

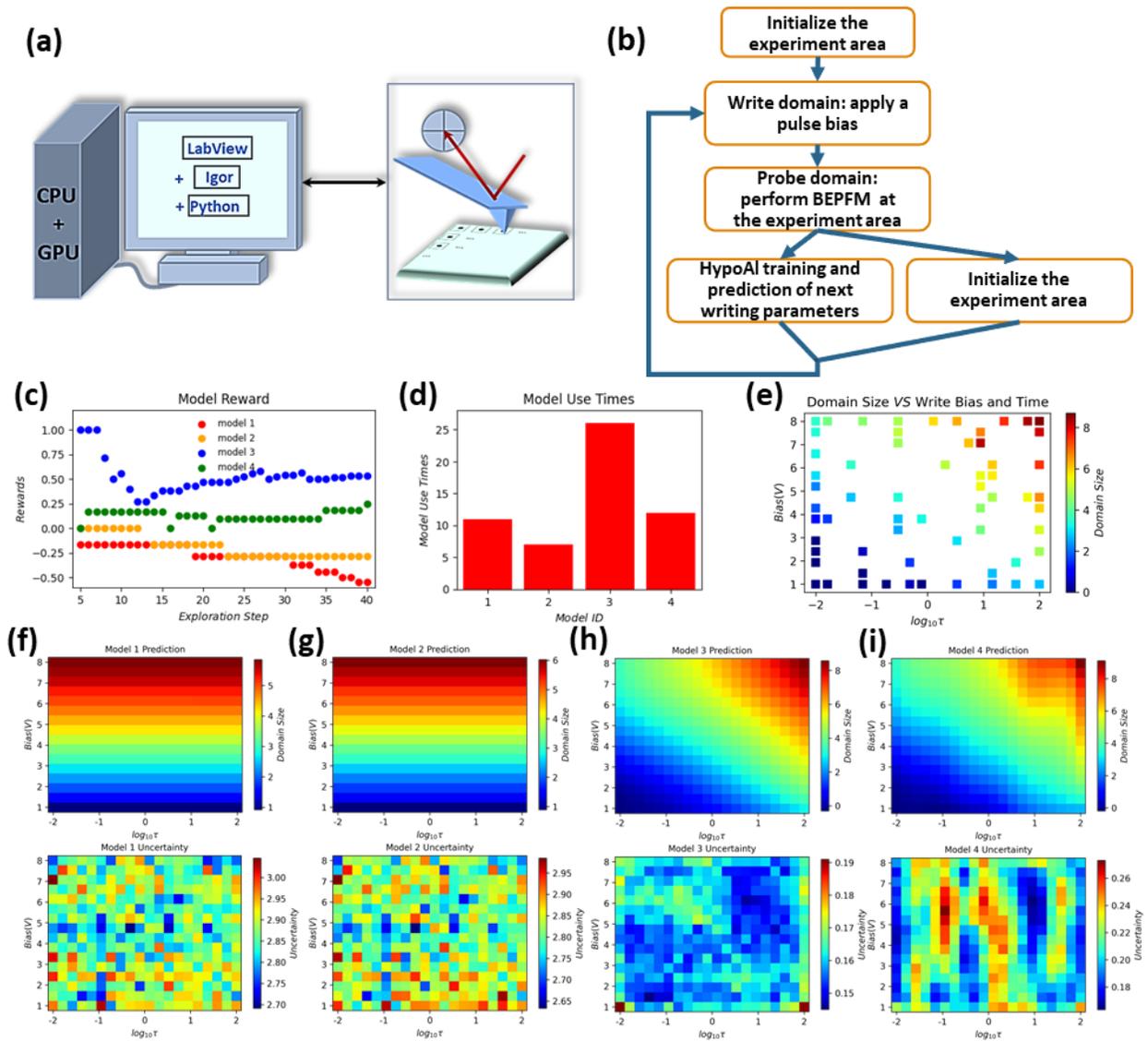

**Figure 10.** Hypothesis learning automated band excitation piezoresponse force microscopy. (a) a schematic showing the integrated system for hypothesis learning automated BEPFM; (b) hypothesis-learning-based automated BEPFM workflow; (c-e) Hypothesis learning BEPFM results of a BaTiO$_3$ (BTO) thin film, (c), model rewards during the hypotheses learning after the initial 5-step warm-up phase during which all models were evaluated at each step. Model III gained a much larger reward than other models, and its reward gradually increased at the latter part of the experiment; (d) model selection in hypotheses-learning, in which model III was selected more often than other models. (e) experimental results of domain size as a function of writing parameters; (f-g) Predictions by all models on the final set of discovered parameters after the competition of the experiment, prediction (top row) and corresponding uncertainties (bottom row) by four different models, respectively. Adapted with permission under a Creative Commons Attribution 4.0 International (CC BY 4.0) from Ref[106], Copyright 2022 ArXiv.

The hypothesis learning based automated experiment is illustrated in Figure 10.[106] Here, the microscope images the selected region of the sample surface and polarizes it into the single-domain state by scanning with the biased tip. With this, the tip is positioned at the center of the image and the bias pulse of opposite polarity with a certain magnitude and duration is applied. After the application of the pulse, the surface is rescanned, and the size of the formed domain is determined. Based on the measured domain size, the parameters for the next pulse are selected and the process is repeated in a fully automated manner.

Initially, 5 steps are performed to yield the initial data for the algorithm. With this in hand, the hypothesis learning algorithm is activated and the evolution of the rewards for different models as a function of experimental step is shown in Figure 10 (c). Here, models 1 and 2 are thermodynamic for two limiting cases of thin and thick domain wall, and do not have time dependence of the domain size. Model 4 is the charge-injection limited model, which consistently gives better agreement with the experiment. However, Model 3 based on the domain wall pinning process yields best predictability. The experimental flow in Figure 10 (d) illustrates how the preferred model is discovered during the experiment. The resultant sampling of the two-dimensional (V, $\tau$) space is shown in Figure 10 (e). Note the preferential sampling close to the edges of the defined parameter space.

Finally, the output of the hypothesis learning experiment is the chosen model and its parameters, i.e., physical law that describe the domain growth in ferroelectric. Here, it is discovered in a fully automated fashion. Notably, the hypothesis learning also allows establishing how the alternative models would describe the data and generate the associated uncertainties. However, it is important to note that the discovery path depends on the chosen reward and model selection during the experiment.

## 8. Some general considerations

To summarize, here we discuss the general framework for the application of the Bayesian active learning methods in scanning probe microscopy, using the example of Piezoresponse Force Microscopy. We pose that any automated experiment workflow requires defining the prior knowledge, the data available during the experiment, and the reward defining the experiment goal. We define the direct and inverse automated experiment tasks, namely the sampling of the a priori known objects in the image plane vs. discovery of the microstructural elements that possess the behaviors of interest. We illustrate the basic principle and formalism behind the simple Gaussian Process models, and illustrate their application for image reconstruction (i.e., sampling sequence of the image or spectral space is predetermined) and in automated experiment. In the latter case, the results of the prior experiment are used to select the locations for the next one.

We further illustrate that simple GP algorithms as applied to imaging or spectral imaging (with defined scalar parameter of interest) offer only marginal improvement compared to the grid sampling methods, with the modest gains in sampling efficiency but greatly increased requirements for compute and drift correction. The reason for this is that the spatially-complex data characteristic for physical imaging cannot be well-represented via the kernel function, and discovery of the details at a certain length scale tends to erase details on a different length scales. These behaviors can be intuitively understood on the full ground truth data, assuming that kernel is roughly equivalent to the correlation function.

**Conclusions**

Comparatively, the Bayesian methods offer very significant advantages when incorporated as a part of structure-property discovery framework. In this, rather than learning the unknown black box function (e.g. image via sequence of pixel by pixel measurements), the automated experiment aims to reconstruct the structure property relationships, e.g. relationship between local structure and local spectra. Here, the full image is available to the algorithm as a prior data. This approach is realized via deep kernel learning, combining the expressive power of deep convolutional networks and flexibility of GP. This approach allows effective reconstruction at ~1% of the imaging budget.

We further discuss the use of Bayesian methods for functional fits, allowing to incorporate the prior information in the form of model selection and prior distribution of model parameters. While more computationally intensive then classical least square fits, this Bayesian inference approach allows significant flexibility for data analysis and allows to introduce the prior knowledge on system behavior and materials properties in principled and reproducible way. We also note that the characteristic aspect of Bayesian methods – namely reliance on priors – that has long been perceived to be its weakness from statistical point of view, in fact make it exceptionally powerful from the physics domain perspective.

Finally, we discuss the structured Gaussian Processes and hypothesis learning as strategies for active experiment that combine the prior physical knowledge of the physical model and flexibility of the pure data-driven Gaussian Processes and demonstrate the fully automated PFM experiment discovering the correct physics behind the domain growth in ferroelectric materials.

While this review is centered around a single scanning probe microscopy method – Piezoresponse Force Microscopy – these principles, strategies, and codes can be directly extended to other imaging methods that allow for structural, functional, and spectroscopic imaging, and combinations of physical methods. For example, the scanning tunneling spectroscopy in scanning tunneling microscopy or force distance curve measurements in conventional atomic force microscopy will follow the same logic. Similar cases will apply to other scanning probe techniques, for example electron energy loss spectroscopy in scanning transmission electron microscopy or optical and mass-spectrometry based chemical imaging. Finally, the same approach will apply for the combination of different techniques, e.g., SPM-scanning electron microscopy, optical microscopy and nanoindentation, or X-ray based local diffraction measurements and SPM or optical imaging.

Finally, we note that implementation of the automated experiments workflows requires three well defined components. One is the engineering controls, allowing the ML algorithms to guide experimental pattern. The second is the ML algorithms that define the experimental path, either in a purely data-driven fashion (GP) or utilizing prior experimental (DKL) or physical (structured GP, hypothesis learning) knowledge. However, the third and key component is the reward structure, or the goal of experiment – either discovery of specific behaviors, general curiosity learning, or some combination of exploration and exploitation, and reflected in the a priori defined features of interest in multidimensional data. With these three components explicitly defined, the instantiation of automated experiments in SPM, STEM, optical, focused X-Ray and other physical and chemical imaging methods becomes straightforward and is ushering in an era of scientific discovery.

**Acknowledgements:**

The conceptualization of the review and the writing of introduction, section 1, 2, 4 was supported by the center for 3D Ferroelectric Microelectronics (3DFeM), an Energy Frontier Research Center funded by the U.S. Department of Energy (DOE), Office of Science, Basic Energy Sciences under Award Number DE-SC0021118. The writing of section 3, 5, and 7 was supported by the Center for Nanophase Materials Sciences (CNMS), which is a U.S. Department of Energy, Office of Science User Facility at Oak Ridge National Laboratory. The scanning probe work with model selection was sponsored by the U.S. Department of Energy (DOE), Office of Basic Energy Sciences, Division of Materials Sciences and Engineering.

**Vocabulary:**

*Active learning:* An approach where a machine learning model interacts with the data generation process updating its knowledge about the system after each new measurement. Experimental pathways (such as selecting the next measurement locations) are automatically adjusted based on this evolving knowledge.

*Gaussian process:* A method for reconstructing an unknown ("black-box") function with quantified uncertainty over a low-dimensional parameter space from sparse measurements.

*Deep kernel learning:* A hybrid of deep neural network and Gaussian process that establishes a correlative relationship between observed structural features and functional property of interest.

*Bayesian inference:* A method for updating probability distributions over parameters of a physical model as more information becomes available.

*Hypothesis learning:* An active learning method for reconstructing the overall data distribution within a small number of measurements and simultaneously learning a correct model of the system's behavior from the list of possible probabilistic models (hypotheses).

**Methods:**

The codes for applying the described tools to experimental data together with the examples are available without restrictions via GitHub:
- Basic GP and BO: https://github.com/ziatdinovmax/GPim
- Deep kernel learning, structured GP, hypothesis learning: https://github.com/ziatdinovmax/gpax

**Notes:**

The authors declare no competing financial interest.

**Associated Content:**

A pre-print version of this work is available at ArXiv.[107]

**References**

1. Gerber, C.; Lang, H. P., How the doors to the nanoworld were opened. *Nat. Nanotechnol.* **2006,** *1* (1), 3-5.
2. Garcia, R.; Perez, R., Dynamic atomic force microscopy methods. *Surf. Sci. Rep.* **2002,** *47* (6-8), 197-301.
3. Giridharagopal, R.; Shao, G. Z.; Groves, C.; Ginger, D. S., New SPM techniques for analyzing OPV materials. *Materials Today* **2010,** *13* (9), 50-56.
4. Binnig, G.; Quate, C. F.; Gerber, C., ATOMIC FORCE MICROSCOPE. *Phys. Rev. Lett.* **1986,** *56* (9), 930-933.

For Table of Contents Only

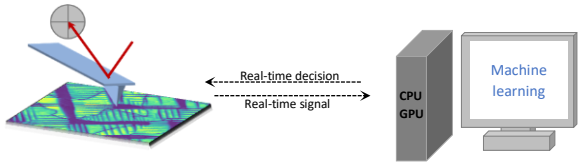